\pdfoutput=1
\documentclass[12pt,a4paper]{article}
\usepackage{jheppub}
\usepackage{amssymb,amsmath,amsfonts}
\usepackage{epsfig}
\usepackage{slashed}

\usepackage{color}

\def\clock{{\count0=\time
           \divide\count0 60
           \ifnum\count0<10 0\fi\the\count0
           \multiply\count0 -60 \advance\count0 \time
           :\ifnum\count0<10 0\fi \the\count0
         }}
\newcommand{\timestamp}{{\small\vbox{\hbox{\tt\jobname.tex}
\hbox{\the\day/\the\month/\the\year, \clock}}}}

\newcommand{\nn}{\nonumber}

\newcommand{\ads}{\mbox{AdS}}
\newcommand{\cft}{\mbox{CFT}}

\newcommand{\be}{\begin{eqnarray}}
\newcommand{\ee}{\end{eqnarray}}
\newcommand{\beqn}{\begin{eqnarray}}
\newcommand{\eeqn}{\end{eqnarray}}
\newcommand{\beq}{\begin{equation}}
\newcommand{\eeq}{\end{equation}}

\newcommand{\grp}[1]{\mathrm{#1}}

\newcommand{\grU}{\grp{U}}

\usepackage{color}

\title{Friedel Oscillations in Holographic Metals}
\author[a]{V. Giangreco M. Puletti,}
\author[b]{S. Nowling,}
\author[b,c]{L. Thorlacius,} 
\author[b,c]{and T. Zingg}

\affiliation[a]  {Department of Fundamental Physics,\\
Chalmers University of Technology, 412 96 G\"oteborg, Sweden}

\affiliation[b]  {Nordita\\
Roslagstullsbacken 23,
SE-106 91 Stockholm, Sweden}
\affiliation[c]{University of Iceland, Science Institute\\
Dunhaga 3, IS-107 Reykjavik, Iceland}

\emailAdd{marotta@chalmers.se}
\emailAdd{nowling@nordita.org}
\emailAdd{larus@nordita.org}
\emailAdd{zingg@nordita.org}

\abstract{In this article we study the conditions under which holographic metallic states 
display Friedel oscillations.  We focus on systems where the bulk charge density is not 
hidden behind a black hole horizon.  Understanding holographic Friedel oscillations 
gives a clean way to characterize the boundary system, complementary to probe fermion calculations.  We find that fermions in a ``hard wall" $\ads$ geometry unambiguously 
display Friedel oscillations.  However, similar oscillations are washed out for electron 
stars, suggesting a smeared continuum of Fermi surfaces. }

\rightline{\vbox{\tiny\hbox{NORDITA-2011-89\,, RH-14-2011}}}

\begin{document}
\maketitle
\flushbottom

\setcounter{page}{1}


\section{Introduction\label{sec:Intro}}
Recent years have seen a number of holographic constructions aimed at 
modeling fermions at finite density in 2+1 dimensions 
(see \cite{Hartnoll:2011fn,Sachdev:2011wg,Iqbal:2011ae} for recent reviews). 
A primary motivation has been to find a suitable gravity dual for 
non-Fermi liquids, that is metals whose low energy dynamics differ from 
those of the usual Landau-Fermi liquid. In conventional metals the robustness 
of Fermi liquid theory has been argued in terms of an IR stable Gaussian fixed 
point of the renormalization group governing the low-energy 
excitations of charged fermions 
about a Fermi surface in 3+1 dimensions \cite{Polchinski:1992ed,Shankar:1994ss}. 
If, on the other hand, the physics of the conduction electrons is essentially confined 
to two spatial dimensions this no longer applies. The low energy excitations of charged 
fermions about a Fermi surface in 2+1 dimensions are strongly coupled and perturbative 
field theory cannot be relied upon \cite{Lee:2009xx,Metlitski:2010aa,Metlitski:2010ab}.

In this context, gauge/gravity duality may provide useful new insights and considerable
effort has been put into developing holographic duals of strongly coupled fermion 
dynamics in 2+1 dimensions \cite{Lee:2008xf,Liu:2009dm,Faulkner:2009wj,Cubrovic:2009ye,Denef:2009yy,Hartnoll:2009ns,Faulkner:2010tq,Sachdev:2010um,Hartnoll:2010gu,Cubrovic:2010bf,Hartnoll:2010xj,Iizuka:2011hg,Hartnoll:2011dm,Arsiwalla:2010bt,Cubrovic:2011xm,Edalati:2011yv,Sachdev:2011ze,Puletti:2010de,Hartnoll:2010ik}. 
An important step is to demonstrate how a Fermi 
surface may be encoded in a holographic geometry. One approach is to couple a 
probe fermion to the bulk gravitational sector and compute the resulting spectral 
density \cite{Lee:2008xf,Liu:2009dm,Cubrovic:2009ye,Faulkner:2009wj,Hartnoll:2011dm}.
This has led to many interesting results, including signals of Fermi surfaces, 
Fermi liquids, and even
non-Fermi liquid behavior.  Interesting characteristics of the probe fermion
arise because the spacetime geometry and other bulk fields serve to provide a 
strongly coupled gapless sector to which the probe is weakly 
coupled \cite{Faulkner:2010tq}.  The probe fermion results are, however, 
rather indirect from the point of view of the holographic background which, 
by definition, does not know about the presence of probes.

In the present paper, we instead look for signals of 
Fermi surfaces in static susceptibilities, such as a current-current correlator at 
finite momentum and zero frequency \cite{Kulaxizi:2008jx}. Such correlation 
functions are inherent to 
the holographic setup and the relevant gauge couplings are determined by
the same bulk Lagrangian as the gravitational background itself.
Specifically, we expect to see non-analytic features, so called Friedel oscillations,
at twice the Fermi momentum in zero temperature static current-current correlation 
functions when there are a discrete set of Fermi surfaces. Friedel oscillations are 
expected whenever there is a sharp Fermi surface, and do not rely upon the 
assumptions of Fermi liquid theory.

The holographic models that we consider have three key ingredients in common. 
One is that the gravitational dual spacetime is asymptotically AdS, which corresponds 
to an underlying conformal symmetry in the boundary field 
theory. Relaxing the condition of conformal symmetry is of considerable interest but 
this will not be pursued here. Another shared ingredient is a $\grU(1)$ electric flux 
threading the asymptotic AdS region, which is the holographic manifestation of a 
non-vanishing chemical potential and charge density in the boundary theory. 
 The finite charge densities at zero temperature indicate that the dual field theory 
 also possesses Fermi surfaces by a generalized Luttinger's 
 theorem \cite{SachdevGenLutt}.  Finally, we have chosen to focus on models 
 where the asymptotic electric flux is sourced by a finite density of charged 
 fermions inside the bulk rather than emanating from behind the event horizon 
 of a charged black hole.

One such system is an electron star, first considered 
in \cite{Hartnoll:2009ns,Hartnoll:2010gu}.  
There the approach is to increase the charge density of bulk fermions, such 
that they may be considered as a fluid of degenerate fermions.  In this limit, 
there is a continuum of bulk Fermi surfaces, distributed radially. Further, in a 
certain parameter regime, one can ignore the effect of gravity on the fluid's local 
equation of state. By setting up the corresponding Oppenheimer-Volkov equations 
it is possible to numerically determine the spacetime's geometry.  Electron stars are 
non-singular geometries possessing Lifshitz scaling regimes in the deep interior.  
One of the initial hopes for electron stars was that, by working with macroscopic 
bulk charge densities, they would make fermionic properties visible at the classical 
level.  Indeed, in \cite{Hartnoll:2011dm, Cubrovic:2011xm}, it was shown, at the probe 
fermion level, that electron stars support many discrete Fermi surfaces which satisfy a 
Luttinger count.  On the other hand, as discussed in \cite{Hartnoll:2010xj}, most of 
the bulk Fermi surfaces do not contribute to de~Haas-Van~Alphen oscillations in a
background magnetic field. These are dominated by a single Fermi surface while
the contributions from the remaining local Fermi surfaces in the bulk fluid wash out. 
We will find similar smearing when we look for Friedel oscillations in electron star 
backgrounds except in this case the signal is completely smeared out. 
  
Another model for holographic Fermi surfaces, which we will refer to as an ``$\ads$ 
hard wall,"  was introduced in \cite{Sachdev:2011ze}. This model was intended as a 
toy model where a finite number of discrete Fermi surfaces could be made manifest.  
Interpreting the difficulties encountered with extremal black holes and electron stars 
as a signal of too many IR degrees of freedom, \cite{Sachdev:2011ze} instead 
works within a confining dual gauge theory, achieved via a truncation of an $\ads$ 
geometry by introducing ``hard wall'' boundary conditions at a finite value 
of the radial variable. The profile of a bulk gauge field in 
the presence of a filled bulk Fermi surface is solved for numerically, neglecting
any gravitational back reaction, and then the effect of the resulting gauge field on 
the bulk fermions is solved for. These steps are iterated towards a self-consistent 
configuration of the fermions and gauge field. A Luttinger count indicated that the 
resulting theory could describe either Fermi or non-Fermi liquids depending on the 
gauge field's boundary conditions at the hard wall \cite{Sachdev:2011ze}. Morally, this 
configuration is similar to that of an electron star with a non-vanishing charge density 
carried by bulk fermions, but differs crucially in the number of IR degrees of freedom. 
Also, the character of bulk Fermi surfaces is very different than in the electron star 
geometry. In the hard wall construction, a bulk Fermi surface is not a local concept, 
but instead each Fermi surface is assigned a radial mode number.  
The non-locality in the bulk Fermi surface leads to a certain factorization in the way 
radial and boundary information is encoded, implying that the AdS space outside the
hard wall acts more or less like a finite size box, as discussed in \cite{Sachdev:2011ze}. 

Our understanding of the application of holography to fermionic systems at finite 
density is in its infancy.  The electron star and hard wall geometries provide two 
systems through which we may develop our intuition.  By contrasting the two 
backgrounds we compare the effects of bulk Fermi surfaces which are local 
{\it vs.} non-local and continuous {\it vs.} discrete.  
In order to see signatures of 
Fermi surfaces in current correlation functions, which involve local bosonic operators, it is necessary to include the effects of bulk fermionic loops. This is because
fermionic operators do not have expectation values.
The primary result of this paper 
will be to demonstrate that bulk Fermi surfaces in an $\ads$ hard wall geometry 
can induce boundary Friedel oscillations, while a similar calculation for electron 
stars shows that the oscillations are washed out due to the continuum of bulk Fermi 
surfaces. These results suggest that discrete bulk Fermi surfaces which are non-local 
in the radial direction induce sharp, discrete boundary Fermi surfaces.

The paper is outlined as follows. We begin with a review of Friedel oscillations 
in a $2+1$ dimensional perturbative system in Section~\ref{sec:Friedel}.  
In Section~\ref{sec:AdSBoxFriedel} we briefly discuss the $\ads$ hard wall 
background introduced in \cite{Sachdev:2011ze} and then we identify the 
source of Friedel oscillations. 
In Section~\ref{sec:ECFriedel} we set up an analogous calculation in an electron 
star geometry but find that there are no boundary Friedel oscillations due to the 
continuum of bulk Fermi surfaces.  Finally, we conclude with a discussion in 
Section~\ref{sec:Discussion}. In order to streamline the presentation the technical 
details are relegated to appendices when possible.


\section{Friedel Oscillations}\label{sec:Friedel}
In systems with sharp Fermi surfaces, zero temperature, static response functions display spatial oscillatory behavior, called Friedel oscillations.%
\footnote{Strictly speaking Friedel oscillations refer to spatial 
oscillations occurring in charge density perturbations. In this article we will use the
term more generally to refer to spatial oscillations in generic transport coefficients  
due to a sharp Fermi surface.} 
Heuristically, this may be understood as follows: Having a sharp, non-analytic, cut in the quasi-particle density 
of states leads to non-analytic features in momentum space static response functions.  
Upon Fourier transformation, these non-analyticities give rise to oscillatory spatial 
features at wavenumber $2 k_f$.\footnote{A simple example is given by the Fourier 
transform of a one-dimensional sharp bump function in momentum space of width 
$w$ and unit height, \beq F.T.[B(k)] = \sin(w x)/w.\eeq}  
Such oscillations occur in all known models with 
sharp Fermi surfaces, for essentially any static response function considered 
({\it i.e.} density-density, spin-spin, {\it etc.}).  Heuristically, the character of response 
functions changes depending on whether the exchanged momentum ``fits" in the 
Fermi sphere. In this section we will outline how this works for $2+1$ dimensional 
non-relativistic fermions, leaving the details for Appendix \ref{ap:NRPol}. 
Understanding the perturbative calculation will be useful to help set notation and 
because it has close parallels with the holographic calculation.  Similar effects 
are expected in non-Fermi liquids since it is the sharpness of the Fermi surface that matters 
rather than the detailed dynamics.  


\subsection{Perturbative Friedel Oscillations in $2+1$ Dimensions\label{sec:NRpol}}

Friedel oscillations are easily visible in the density-density polarization ({\it i.e.} 
susceptibility), denoted by $\Pi$.  For simplicity we will focus on a system of spinless 
fermions with a Fermi momentum, $k_f$. For weakly coupled fermions this is textbook 
material, see for example  \cite{Fetter,Kapusta:2006pm}.   

In a unitary system we can use a Lehmann representation to capture the full frequency 
dependence of two point functions.  The time ordered Green's function for free fermions 
in a translationally and rotationally invariant system takes the form 
\beqn
\label{green_function_ex}
\mathcal{G}(k,\omega) &=&  \frac{\theta(|k|-k_f)}{\omega-E_k + i\eta } 
+ \frac{\theta(k_f-|k|)}{\omega-E_k - i\eta} \,,
\eeqn
in $(\omega,k)$ space.  Using the fermion's Green's function, an application of Kubo's formula indicates that the leading contribution to the polarization is   
\beq
\Pi(q,\nu) = -i e^2 \int \frac{d\omega d\vec{k}}{(2\pi)^{D+1}}\mathcal{G} (k,\omega)\mathcal{G} 
(k+q,\omega+\nu)\,.
\eeq


In the static limit the polarization tensor is real
\beqn
\mathrm{Im}\ \Pi(q,0)&=&0\, ,\\
\mathrm{Re}\ \Pi(q,0) &=&-\frac{e^2}{2\pi^2}\mathcal{P}\int d k 
d\theta\ \theta(k_f-k)\frac{k}{E_{k+q}-E_{k}}\,.
\label{eq:NRGeneralPol}
\eeqn
For spherical Fermi surfaces the integrand is singular when $q=2k_f$.  
Even without knowing the precise form of the dispersion relation, the 
polarization tensor is non-analytic in momentum space and thus has
oscillatory behavior in position space.

\begin{figure}[height=0cm]
\begin{center}
\includegraphics{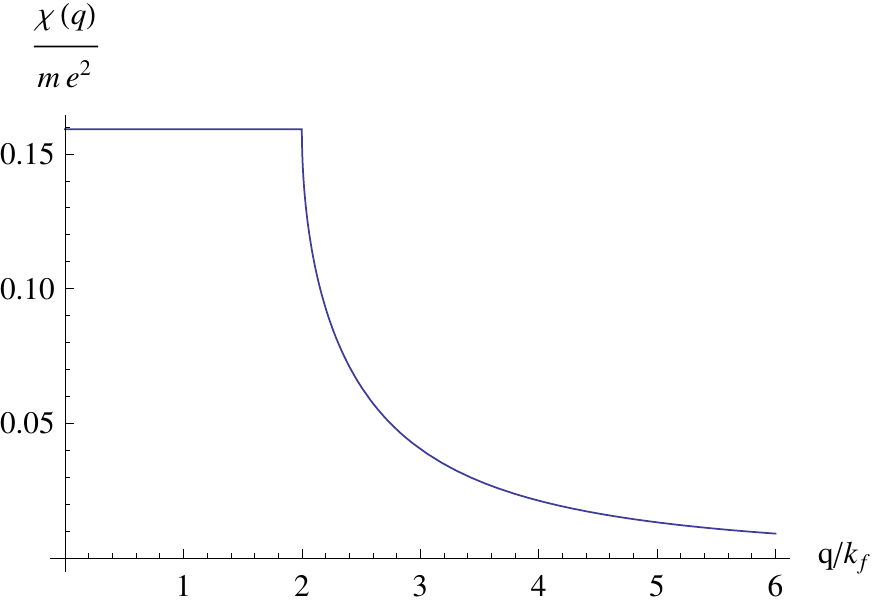}
\caption{The non-relativistic static susceptibility, $\chi(q)= -\Pi(q),$ in $2+1$ dimensions 
for spinless fermions of mass $m$ and charge $e$. \label{fig:NRFriedel}}
\end{center}
\end{figure}

For concreteness, consider non-relativistic free fermions with $E_{k}=\frac{k^2}{2m}$.
In this case (\ref{eq:NRGeneralPol}) reduces to 
\beqn
 \Pi(q,0) 
&=& -\frac{e^2m}{2 \pi^2 q}\mathcal{P}\int d k d\theta\ \theta(k_f-k)\frac{k}{q/2
+k\cos(\theta)}\\
&=&-\frac{e^2m}{2\pi}\left(1-\sqrt{1-\left(\frac{2k_f}{q}\right)^2}\theta(q-2k_f)\right)\,.
\eeqn 

The Kubo formula relates the charge density susceptibility to the polarization tensor
\beqn
\chi(q)\equiv \langle \rho(-q)\rho(q)\rangle_R = -\Pi_R(q)=-\Pi(q)\,.
\eeqn 
In the last equality we have used the fact that the static polarization tensor is real to 
equate its time-ordered and retarded incarnations.  In momentum space the 
non-analyticity at $q=2k_f$ is clearly visible in Figure~\ref{fig:NRFriedel}.


\section{Friedel Oscillations in an $\ads$ Hard Wall Geometry}
\label{sec:AdSBoxFriedel}

As mentioned in Section~\ref{sec:Intro}, a number of holographic models have 
been introduced to describe Fermi surfaces at strong coupling. To reduce the 
number of IR degrees of freedom, the author of \cite{Sachdev:2011ze} proposed to 
work with confining boundary 
field theories.  As is familiar from using holography to study QCD, we can model 
confinement with a hard wall placed in the bulk \cite{Erlich:2005qh}.  
Although this model does not include the effects of a bulk Fermi surface on the metric 
degrees of freedom, the Fermi surface does back react onto the gauge field's profile.  
Therefore, in this model one can study the effects of a bulk Fermi surface outside of a probe 
limit on fermionic or current correlators, but not on stress tensor correlators. 
Here we will focus on current correlators and show that they display boundary Friedel 
oscillations. 


\subsection{Preliminaries}
\label{sec:preliminaries_AdSBox}

Following~\cite{Sachdev:2011ze}, we begin with pure $\ads_4$, 
\beq 
\label{metric_ads4}
ds^2 = \frac{L^2}{z^2}\left(-dt^2+dz^2+d\vec{x}^2\right)\,.
\eeq  
In order to control the number of fields in the dual field theory's IR limit, one introduces 
a hard wall cutoff, $z_m$.  We can imagine that the cutoff is an approximation to some 
more interesting geometric effect~\cite{Witten:1998zw}. On this background geometry 
we place both fermions and gauge fields governed by the standard action 
\beq
\label{eq:SachAction}
S = \int d^4 x \sqrt{-g}\left[\frac{1}{4 e^2} F^2 + \overline{\Psi}\left(i\slashed D 
+ m\right)\Psi \right].
\eeq  
The covariant derivative is 
$D_\mu = \partial_\mu +\frac{1}{4}\omega_{a b \mu}\Gamma^{a b} -i q A_\mu\, .$
We will be interested in constructing the background gauge field profiles and 
understanding linear response theory in the gauge sector on large length scales.  
For these tasks it is sufficient to integrate out the fermions and use the resulting 
low energy effective action
\beq 
S =   \int d^4 x \sqrt{-g}\frac{1}{4 e^2} F^2 
+\mathrm{Tr}\ln \left[  i\slashed D +  m\right]\,.
\eeq 
In addition to the differential equations coming from Eq.~\eqref{eq:SachAction}, 
one must also specify boundary conditions at the hard wall.  
Because the geometry is regular, it is natural to impose self-adjoint boundary 
conditions on the fermions~\cite{Sachdev:2011ze}. Such boundary conditions 
are quite strong, however, ruling out any dissipation in the fermionic sector in 
the large $N$ limit.

With the boundary conditions in place, we seek to fill fermion states up to the 
ground state energy as well as determining the self-consistent gauge field profile.  
In practice this may be done through an iterative process. As discussed 
in~\cite{Sachdev:2011ze}, 
Neumann boundary 
conditions for $A_0$ at the hard wall, are a necessary condition to have a boundary Fermi liquid in this set up.  Any other 
self-adjoint boundary condition leads to non-Fermi liquid behavior. 
The resulting background consists of a filled sea of fermion single particle 
wave functions, each indexed by boundary momentum as well as a radial 
mode number.  In addition one must include the backreaction of the Fermi 
sea on the gauge field.  For chemical potentials above the gap, the system 
has compressible \cite{Huijse:2011hp} gapless excitations and a sharp Fermi 
surface.  

It is useful to consider the time ordered Green's function of these normalizable 
fermions. Using a Lehmann representation we have
\beqn\label{eq:FermGF}
G(\omega,k,z,z') &=& \sum_{\ell\neq 0}  \left(\frac{1}{\omega-E_\ell(k)
+i\, \eta\, \mathrm{sign}(E_\ell(k))}\right)\chi_{\ell,k}(z)\chi^\dagger_{\ell,k}(z')\,,
\eeqn
where the wave functions and energies are those numerically determined when 
solving for the background, and $E_\ell(k)$ is the eigenvalue of the Dirac 
Hamiltonian with spatial momentum $k$ and radial mode number $\ell.$   


\subsection{Mean Field Susceptibility\label{sec:MFBoxSus}}

Here we turn to the problem of linear response theory in hard wall $\ads$ 
in the radial gauge, $A_z=0$. The (static) classical wave equation for the 
potential is 
\beq 
\left(\partial_z^2-k_x^2\right)A_0(k_x,z)=0\,.
\eeq
For definiteness we have taken the momentum to be in the $x$ direction 
and we focus on a gauge invariant sector of perturbations $A_0(k_x)$.  
Since the hard wall is a regular point in the geometry, it is natural to 
impose the same boundary conditions that are used in constructing the 
background.  We can find the classical bulk to boundary Green's function 
({\it i.e.} the solution which goes to one at the boundary) analytically.
\beqn\label{eq:AdSBoxLeading}
G^{B\partial}_{0}(k_x,z) &=& \cosh(k_x z)-\tanh(k_xz_m)\sinh(k_xz)\,,\\
\rightarrow \langle \rho(-k_x)\rho(k_x)\rangle_{0} 
&=& -k_x \tanh(k_x z_m)\eeqn
As expected, the classical contribution to the susceptibility does not know about 
the Fermi surface.  We must go away from mean field theory and consider 
virtual particle-hole exchanges.


\subsection{Polarization Corrections\label{sec:BoxVacPol}}
In order to proceed, let us imagine integrating out the fermions in the bulk.  
Graphically, the first correction to the gauge field's effective action comes 
from a bubble diagram with fermions in the loop.\footnote{At zero momentum 
the effective action for $A_0$ also has a tadpole contribution analogous to 
the one used to determine the gauge field's background profile. This term 
does not affect our results below.}  If we expand out the fermion's functional 
determinant to second order in gauge fields, we have a term of the form 
\beq 
\label{spol_adsbox}
S_{Pol} = \frac{ e^2  }{2}\int dz dz' d\vec{x}\ 
A_\mu(z,-\vec{k})\Pi^{\mu\nu}(z,z',\vec{k})A_\nu(z',\vec{k})\,.
\eeq  
The one loop vacuum polarization tensor is written
\beq
\Pi^{\mu\nu}(z,z',\nu,\vec{k}) = - \int \frac{d\omega d^{2}p}{(2\pi)^3}  
\mathrm{Tr}\ \left[M^\mu G(\omega,p,z,z')M^\nu  
G(\omega+\nu,k+p,z',z)\right]\,,
\eeq
where $M^\mu=-q \Gamma^0\Gamma^\mu.$  
The reason that $M^\mu\neq-q \Gamma^\mu$ is that we have chosen to 
work with the Green's function involving $\chi_{k,\ell}(z)\chi_{k,\ell}^\dagger(z')$ 
rather than $\chi_{k,\ell}(z)\overline{\chi}_{k,\ell}(z')$.  In what follows we will limit 
our discussion to the static polarization tensor, 
$\Pi^{\mu\nu}(z,z',0,\vec{p})\equiv \Pi^{\mu\nu}(z,z',\vec{p})$.
In the background geometry there are three parameters, namely $\mu\,, q\,, e$, 
and only two of them are fixed by the classical geometry~\cite{Sachdev:2011ze}. 
The relative size of the loop correction 
\eqref{spol_adsbox} is determined by the third one.\footnote{To anticipate the 
notation used in Section~\ref{sec:ECFriedel} below, 
we have rescaled the gauge field $A_\mu\rightarrow {eL\over \kappa} A_\mu$, as 
in Eq. \eqref{gauge_field_EC}, with $L$ the curvature radius, and $\kappa$ the 
Newton constant. With this normalization and with the metric as in 
Eq. \eqref{metric_ads4}, the background action has an overall factor of 
${L^2\over \kappa^2}$, which we have factored out in \eqref{spol_adsbox}.}

In general, evaluating the vacuum polarization in curved space is a rather difficult
problem.  There are the usual ultraviolet divergences occurring due to large $\ell$ 
and $k$.  Also, one might worry about the cornucopia of fields predicted by string theory 
predict in an eventual top down approach. As discussed in \cite{Kapusta:2006pm}, 
we can split the vacuum polarization into two pieces: the (uninteresting) contributions 
present in vacuum and the relative polarization due to a chemical potential,
\beq \Pi^{\mu\nu}=\Pi^{\mu\nu}|_{vac}+\Pi_{rel}^{\mu\nu}\,.\eeq  
If the chemical potential is not too high, we can assume that the vacuum part only 
contributes to the ultraviolet renormalization of parameters in  Eq. (\ref{eq:SachAction}), 
and is not relevant to the long wavelength physics associated with Friedel oscillations.

If we denote the vacuum time-ordered bulk fermion Green's function by $G^0$, the 
relative polarization tensor may be written as,
\beqn\label{eq:RelPol}
\Pi^{\mu\nu}_{rel}(z,z',\vec{k})
&=&- \int \frac{d\omega \, d^{2}p}{(2\pi)^3}  \mathrm{Tr}
\left[M^\mu G(\omega,\vec{p},z,z')M^\nu  G(\omega ,\vec{k}+\vec{p},z',z)\right]
\nonumber\\
&&\qquad \qquad  - \mathrm{Tr}\left[M^\mu G^0(\omega,\vec{p},z,z')M^\nu  
G^0(\omega ,\vec{k}+\vec{p},z',z)\right]\,.
\eeqn
We analyze the relative polarization in Appendix \ref{ap:HolRelPol}. The full result 
is in general quite involved but fortunately we do not need it in full detail.
Let us list some important features. First of all, the relative polarization is finite in 
the bulk UV. This is analogous to the usual finite temperature field theory statement 
that the only UV divergences are vacuum divergences. Second, the integrand cleanly 
splits into terms unrelated to the Fermi surface which are analytic in the external 
momenta near $2k_f$ and contributions that have singular points at $k=2 k_f$.  
These singularities do not lead to actual divergences in the polarization but they 
are the source of bulk Friedel oscillations in much the same way as in the 
non-relativistic example in Section~\ref{sec:NRpol}. 

\subsubsection{Illustrative Limits}

As discussed in detail in Appendix \ref{ap:Simplification}, there is a limit where the 
relative polarization radically simplifies and one can obtain an (almost) analytic 
expression.  If we consider a high confinement scale (small $z_m$) then all 
contributions to the loop diagram with non-zero relative radial momentum\footnote{More 
precisely contributions where the fermion propagators have different radial mode 
numbers.} are suppressed by factors of $z_m$ and may be dropped. The integrand 
then only receives contributions from near a Fermi surface.

\begin{figure}[height=.1cm]
\begin{center}
\includegraphics{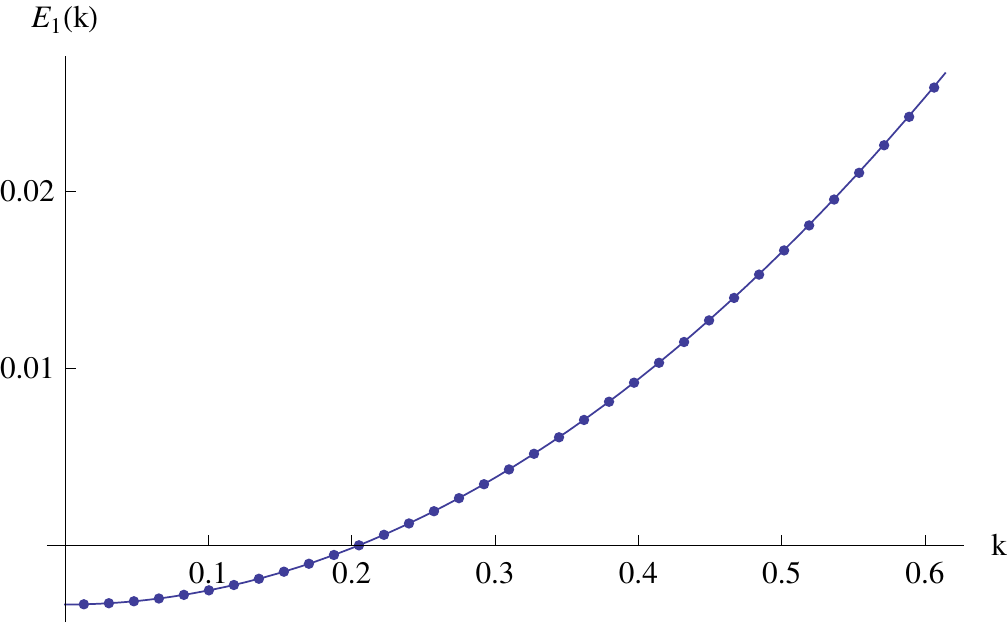}
\caption{The energy dispersion relation for the $\ell=1$ family of eigenvalues 
for $(m,q,e,\mu,z_m)=(1,1,\sqrt{3},6.287,.5)$ with $k_f \sim.20 $ as well as the best 
fit curve $E_1\sim -.003+.8k^2.$  \label{fig:Dispersion}}
\end{center}
\end{figure}

We can make a stronger statement if we take a non-relativistic limit and tune the 
chemical potential towards the fermion's bulk gap energy, leading to a Fermi surface 
of small volume.  For simplicity, consider a background with a single Fermi surface, 
of mode number $1$.  As seen in Figure~\ref{fig:Dispersion}, the numerically determined 
dispersion relation is very well approximated by a quadratic form near the Fermi 
surface
\beq
E_1(k)\sim a_1+b_1 k^2\,. \label{eq:dispersion}
\eeq
Assuming this functional form, the vacuum polarization is well approximated by 
\beqn\ \Pi_{rel}^{\mu\nu}(z,z',\vec{k})&=& -
\lambda \mathrm{Tr} \left(M^\mu\ \chi_{1,0 }(z)\chi_{1,0 }^\dagger(z')
M^\nu\ \chi_{1, \vec{k} }(z')\chi_{1, \vec{k}}^\dagger(z)\right) \nonumber\\
&&\times\left(1-\sqrt{1-\left(\frac{2k_f}{|\vec{k}|}\right)^2}\theta(|\vec{k}|-2k_f)
\right)\,, \label{eq:AdSBoxPolNR}
\eeqn
as discussed in Appendix \ref{ap:Simplification}.
We have introduced an effective expansion parameter, 
$\lambda \equiv \frac{1}{4\pi b_1}$ which encodes the shape of the spectrum.  
The source of bulk (and boundary) Friedel oscillation is now explicit in this 
non-relativistic limit.  Interestingly, there is an important difference compared to the 
perturbative calculation in Section~\ref{sec:NRpol} in that the polarization is a 
non-diagonal tensor. This amounts to non-vanishing mixed density-current correlation 
functions in the holographic dual.

\subsection{One-Loop Corrected Bulk-Boundary Green Functions}
Returning to the gauge invariant sector, it is straightforward to set up of the 
diffeo-integral equations for the loop corrected gauge field equations of motion,
\beqn
\label{aeq_oneloop}
\left(\partial_z^2-k_x^2\right)A_0(k_x,z)&=&- e^2 
{\int} dz'\left[\Pi_{rel}^{00}(z,z',k_x) A_0(k_x,z')
 +\Pi_{rel}^{0x}(z,z',k_x) A_x(k_x,z')\right] . \nonumber\\ 
 &&\qquad \qquad
\eeqn
The absence of metric factors on the left-hand side is a peculiarity of the gauge field
equation of motion in $\ads_4$.  We will work in a background where the gauge 
non-trivial part of $A_x$ vanishes such that we can drop the contribution from the 
second term on the right hand side of \eqref{aeq_oneloop}.  

If we have a high confinement scale and a small Fermi surface volume, the 
diffeo-integral equation can be solved perturbatively in $\lambda $ for the bulk to 
boundary Green's functions
\beqn
G^{B\partial}(k_x,z) &=& G^{B\partial}_{0}(k_x,z)
+ \lambda  G^{B\partial}_{1}(k_x,z)+....\qquad \ \ (t-\mathrm{component})\,\,.
\eeqn
This expansion should be valid as long as loop effects are small.  
Expanding \eqref{aeq_oneloop} we find,
\begin{figure}[t]
\begin{center}
\includegraphics{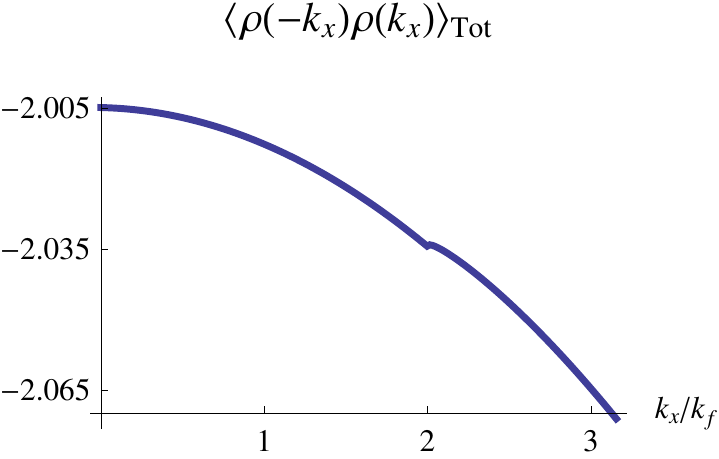}
\includegraphics{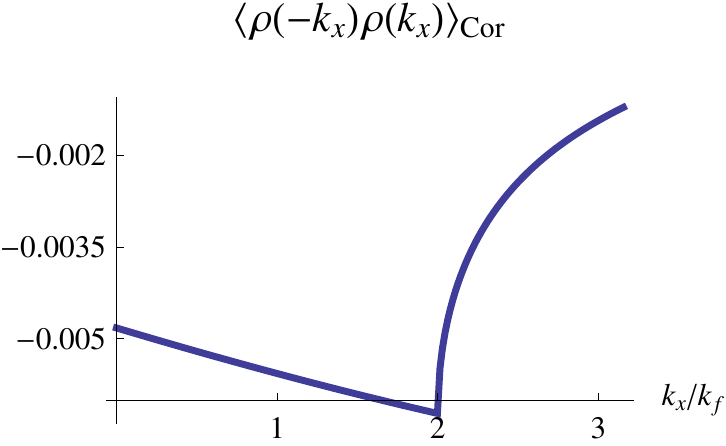}\\
\caption{Correlation functions for $(m,q,e,\mu,z_m)=(1,1,\sqrt{3},6.287,.5)$ with 
$k_f \sim.20 $ in a hard wall geometry. On the left, the full density-density correlation 
function computed by including the one-loop effect induced by the polarization term 
\eqref{spol_adsbox}. On the right, only the one-loop contribution to the 
density-density correlator is plotted. \label{fig:DensDensD}}
\end{center}
\end{figure}
\beqn
0&=& \left( \partial^2_z-k_x^2 \right) G^{B\partial}_{0}(k_x,z)\,,\\
0&=& \left( \partial^2_z-k_x^2 \right) G^{B\partial}_{1}(k_x,z)
+\frac{e^2}{\lambda }\int dz' \Pi_{rel}^{00}(z,z',k_x) 
G^{B\partial}_{0}(k_x,z')\label{eq:DiffInt}\,.
\eeqn
The leading contribution, $G^{B\partial}_0$, was determined in 
Eq. (\ref{eq:AdSBoxLeading}).  Since the leading contribution satisfies the physical 
boundary condition, $G^{B\partial}_0(k,0)=1$, the corrections must satisfy Dirichlet 
conditions near $z\sim0$.  


\subsubsection{Numerical Results\label{sec:Numerics}}

The perturbative equations (\ref{eq:DiffInt}) can be solved numerically and using the 
$\ads/\cft$ dictionary it is then straightforward to translate the bulk to boundary 
Green's functions into correlation functions in the boundary theory. 
Figure~\ref{fig:DensDensD} shows the resulting density-density susceptibility with 
and without the uninteresting classical contribution.  At the $\mathcal{O}(1/N)$ level 
we clearly see a non-analyticity at $k=2k_f$, the hallmark of a sharp Fermi surface.  
The analysis in this section is performed for Neumann hard wall conditions. It would be interesting to extend the analysis 
to other boundary conditions. Intuition from field theory suggests that Friedel oscillations should still be visible whenever there is a sharp Fermi surface regardless of the stability 
of the quasi-particles at the Fermi surface.

The loop corrections involve the effective loop parameter, $ \frac{L}{\kappa},$ 
encoding $1/N$ corrections which did not play a role in when determining the 
classical gauge field background. The value of the loop parameter does not 
affect the position of the non-analyticity in \eqref{aeq_oneloop} and we have set
it to one in order to maximize the amplitude of boundary Friedel oscillations in
our numerical results.


This example illustrates several important points relevant when trying to construct 
boundary Fermi surfaces from Fermi surfaces in the bulk.  Firstly, even when the role 
of the bulk Fermi surface is manifest in the construction of the background profiles, 
classically computed linear response susceptibilities typically do not reveal the surface.  
As in perturbative descriptions, one is looking for evidence of particle hole pairs in 
bosonic observables, which are only present in loops (though the loops are now in 
the bulk).  As discussed in \cite{Sachdev:2011ze}, Luttinger's theorem fixes the 
boundary value of $k_f$ to be the same as the bulk value.  The amplitude of Friedel 
oscillations is thus $N$ dependent, but not the wave vector at which they occur.

Secondly, the presence of a bulk Fermi surface does not guarantee a clear signal in 
the boundary theory.  To be visible, effects of bulk Fermi surfaces must act coherently 
across the radial direction.  For example, when we compute corrected Schwinger-Dyson 
equations, there are convolution integrals in the radial direction.   In order to prevent 
the convolution integrals from washing out the effects of a bulk Fermi surface, the 
spectrum of radial momenta must be gapped.  In the model of \cite{Sachdev:2011ze}, 
it is the confinement scale which introduces the gap.  

Thirdly, even if one can neglect smearing due to convolutions, it is necessary to have 
factorization between the dependence on the boundary momenta and the radial positions.  
In this hard wall example this feature is visible in the form of Eq. (\ref{eq:AdSBoxPolNR}) 
(or more generally in Eq. (\ref{eq:AdSBoxPol})).  The relevant singularities in the integrand 
are only functions of boundary momenta.  In the hard wall example it is the radial 
non-locality of the Fermi surface which makes this happen.  

In the next section we will discuss electron star geometries which also have a finite 
density of fermions in the bulk.   However, this geometry differs in two crucial ways from 
the hard wall model. The spectrum of radial momenta is not gapped and the bulk 
fermionic features (such as the Fermi momenta) are local functions of the radial 
positions.


\section{Absence of Friedel Oscillations in Electron Stars}
\label{sec:ECFriedel}

Another holographic model for the physics of finite density fermions are electron 
stars \cite{Hartnoll:2009ns,Hartnoll:2010gu}.  The basic elements include a fluid of 
charged electrons treated as an ideal fluid of non-interacting particles at zero 
temperature, {\it i.e.} a Thomas-Fermi approximation.  The star geometry carries 
charge more efficiently than an extremal black hole and hence it is thermodynamically 
preferred at zero temperature.  In the deep interior the gravitational solution goes over 
to a Lifshitz scaling solution.  In \cite{Puletti:2010de,Hartnoll:2010ik}, the finite 
temperature extension of this geometry was found.  As for holographic superconductors, 
at low temperatures it is thermodynamically preferred for the charged black holes to 
expel charged matter.  The preferred low temperature geometry consists of a charged 
black hole surrounded by a fermion fluid of the same sign charge.  Above a critical 
temperature there is a third order phase transition from an electron cloud over to 
an $\ads$-RN black hole. We will refer to the zero temperature geometry as an 
electron star and to its finite temperature generalization as an electron cloud.  
Although both the hard wall fermion background and the electron star have bulk Fermi 
surfaces there are important differences, which will allow us to test the lessons 
mentioned in Section~\ref{sec:Numerics}. Ultimately, we find that bulk Friedel 
oscillations will wash out, leaving no sharp signature in the boundary current-current 
correlation function in the electron star.

\subsection{Electron Star Background}
The electron star geometry is given by\footnote{Note that we have slightly 
redefined the metric functions $f$ and $g$ relative to \cite{Hartnoll:2010gu}.  We will 
also denote the dynamical critical exponent by $s$ rather than the usual $z$ to avoid 
confusion with the radial coordinate.}
\beqn 
ds^2 &=& \frac{L^2}{z^2}\left(- f(z) dt^2+g(z) dz^2 + d\vec{x}^2\right)\\
\label{gauge_field_EC}
A &=& {e L\over \kappa} h(z) dt \, ,
\eeqn 
 where $e$ is the fermion charge, $L$ is the curvature radius, and $\kappa$ is Newton's constant. In each local patch there is a degenerate fluid of free fermions modeled as an 
ideal fluid described with velocity, pressure, energy density, and charge density, 
$(u_\mu,\hat{p},\hat{\rho},\hat{\sigma})$.\footnote{The fluid variables are rescaled 
as in~\cite{Hartnoll:2010gu},
\be
\label{def_thermo_u}
p={1\over L^2 \kappa^2 } \hat p\,, 
\qquad
\rho={1\over L^2 \kappa^2} \hat \rho\,, 
\qquad
\sigma={1\over e L^2 \kappa} \hat \sigma\,.
\ee
Furthermore, the mass of the fermions and the constant proportional to the spin in 
the fermion equation of state are rescaled according to $m= {e\over \kappa} \hat m$, 
$\beta= {\kappa^2 \over e^4 L^2} \hat \beta$.}
In the scaling regime the geometry takes the form 
\beqn
f\sim \frac{1}{z^{2(s-1)}}\,,\qquad 
g\sim g_\infty\,,\qquad \mathrm{and}\qquad
h\sim\frac{h_\infty}{z^s}\,.\eeqn  
In the rest of the spacetime the background profiles are non-singular functions of the 
radial coordinate interpolating between an interior Lifshitz region and a near-boundary 
$\ads$ region.

One of the initial motivations for the electron star model was to make the fermionic 
features more dominant by working at finite fermion densities.  
Indeed, in \cite{Hartnoll:2010xj}, it was shown that the thermodynamic potential 
displays de Haas-van Alphen oscillations due to a bulk Fermi surface.  
However it is also clear that the fermionic character of the electron star geometry 
is quite different than in the hard wall model of Section~\ref{sec:AdSBoxFriedel}.
Most importantly, there is a continuum of Fermi surfaces in the Thomas-Fermi 
approximation.  At each radial position there is an effective local chemical potential 
\beq \hat{\mu} = \frac{z h}{\sqrt{f}}\,.\eeq  
A second important difference is the electron star's lack of confinement and a relative 
surplus of infrared degrees of freedom.  The emergence of a Lifshitz scaling 
region in the geometry's interior implies 
that there is no gap in the allowed radial momenta.  

In the rest of this section we discuss Friedel oscillations in the electron cloud geometry.  
We work with the finite temperature electron cloud geometry rather than the zero 
temperature electron star. This is purely for convenience when carrying out numerical 
calculations. Since the electron star geometry, including the Lifshitz scaling region, is 
recovered when the temperature of an electron cloud geometry is taken to be very low compared to the scale set by the chemical potential, we can expect
to recover electron star correlation functions from the ones obtained from  
electron clouds in the low temperature limit. Furthermore, as the temperature of an 
electron cloud geometry is given by the Hawking temperature of the black hole horizon
below the electron cloud while treating the electron cloud itself as a zero temperature
fluid \cite{Puletti:2010de}, any bulk ``medium" effects should be visible even at finite 
temperature. 

As a second technical simplification we will not look for Friedel oscillations in the sound 
channel, the natural generalization of Section~\ref{sec:MFBoxSus}.  Instead, we will use 
the fact that any channel is expected to display Friedel oscillations in the presence of
a sharp Fermi surface and work with the much simpler shear channel.

\subsection{Electron Star Static Correlation Functions: Shear Channel}

Not surprisingly, analyzing the linear response theory for electron stars and electron 
clouds is considerably more complicated than for the hard wall.
If we consider a general perturbation with boundary frequency and momentum 
$(\omega,k_y)$, we can classify the allowed perturbations according to their behavior 
under the residual $\mathbb{Z}_2$ symmetry, ($x\rightarrow -x$) as discussed 
in~\cite{Edalati:2010hk,Edalati:2010pn}. In asymptotically $\ads_4$, there are two 
families of perturbations in the presence of the fluid in the radial gauge 
$\delta A_z=0\,, \delta g_{z\mu}=0$,
\beqn 
\mathrm{odd}:&& (\delta g_{xy}, \delta g_{xt}, \delta A_x, \delta u_x)
\nn\\
\mathrm{even}:&& (\delta g_{xx},\delta g_{yy},\delta g_{tt},\delta g_{ty}, 
\delta A_y, \delta A_t, \delta u_t,\delta u_z,\delta u_y).
\eeqn   
The family of odd perturbations give rise to shear modes while the even family 
corresponds to sound modes in the dual field theory.  Here we will focus on the 
shear channel because it includes a smaller set of fields.  

\subsubsection{Static Limit}
In general, the gravitational response theory is complicated by the fact that there are 
gravitational gauge symmetries in the form of diffeomorphisms.  In the static limit there 
is a simplification and we only need to retain $\delta g^{t}_{\ x},\delta u_x ,$ and 
$\delta A_x $. It is straightforward to set up the classical fluctuation equations 
describing gauge invariant shear modes in the $\omega\rightarrow0$ limit,
\beqn
\label{eq:StaticShearEq}
0&=& (\delta g^{t}_{\ x}) ''+\frac{ (g  \hat{\mu}  \hat{ \sigma}  -4)}{2 z}
(\delta g^{t}_{\ x}) ' -g k_y^2  \delta g^{t}_{\ x} -\frac{2 \sqrt{f } g   
\hat{\sigma} }{z} \hat{\mu}\delta u_x +2 z^2 h'  (\delta A_x)' \,,\\
0&=&(\delta A_x)''-gk_y^2\delta A_x +\frac{ h'}{f}(\delta g^{t}_{\ x}) '
-\left( \frac{g' }{g }+ \frac{g \hat{\mu}  \hat{ \sigma}}{2 z}\right) 
(\delta A_x)'+\frac{g\hat{\sigma}}{z^2}\delta u_x\,.\eeqn
In addition to the equations of motion, the divergence of the stress-energy tensor fixes 
\beq  \delta A_x = -\hat{\mu}\delta u_x\,.\eeq
For an electron cloud geometry, it is straightforward to compute Euclidean boundary susceptibilities by imposing regularity conditions at the horizon.  As in the non-relativistic example we find a real static susceptibility and hence the Euclidean result equals its 
Lorentzian counterpart. 

\begin{figure}[height=.1cm]
\begin{center}
\includegraphics{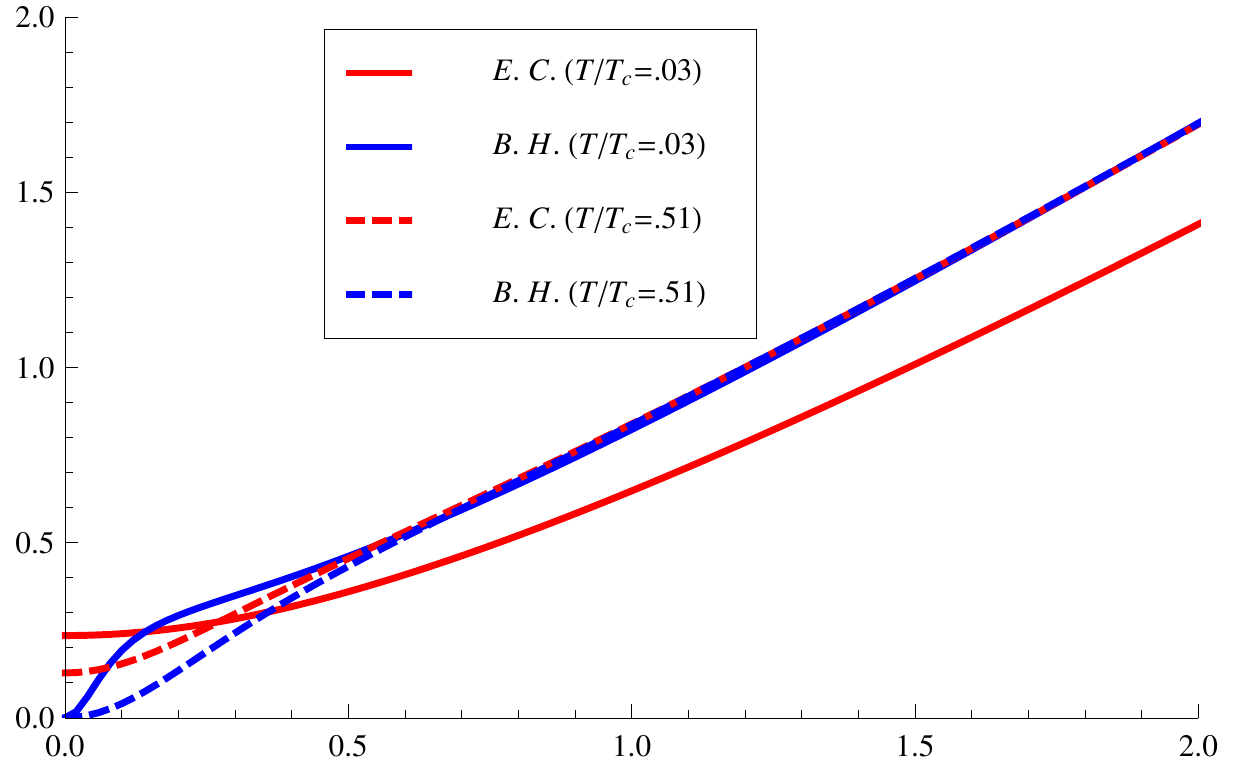}
\caption{Classical contribution to static correlation function in the presence of a fluid 
with parameters $(\hat{\beta} = 10,\hat{m}=\frac{55}{100})$ and for RN black holes at 
moderate and low temperatures in units of $T_c$.  \label{fig:ClassicalFluid}}
\end{center}
\end{figure}

As in the $\ads$ hard wall example we need to consider vacuum polarization effects.
Indeed, neither the differential equation \eqref{eq:StaticShearEq} nor the boundary conditions can know about a Fermi surface. Within classical fluid response, there is nowhere 
for non-analyticity in the boundary momenta to enter the problem, and, accordingly, there is no evidence of a sharp Fermi surface in Figure~\ref{fig:ClassicalFluid}.  A similar result was found
in~\cite{Kulaxizi:2008jx} for the $D4-D8-\overline{D8}$ system.

\subsection{Polarized Electron Star}

We would like to include polarization effects on the shear modes' equations of motion. 
This represents a deviation from the idealized perfect fluid modeling.  Even in flat space 
it is a complicated problem to go from a microscopic field theory description to a long 
wavelength hydrodynamical picture.  In the case at hand we have the added challenge 
that the radial and boundary momenta enter the problem at different levels.  Boundary 
momenta are essentially arbitrary, being fixed by boundary conditions, whereas bulk 
momenta are set by the scale of variation in the bulk space time.  Here we will only be 
interested in effects which lead to non-analyticity and bulk Friedel oscillations, and will 
ignore other, possibly larger, effects of interactions that only contribute analytic terms to
the polarization tensor.

Noting that the vacuum polarization is a  {\it short distance} effect when compared to the 
scale of variation in the radial direction, we will introduce a poor man's modeling of the polarization,
\beqn 
S_{Pol}= \frac{e^2}{2}\int dz dz' d^{2}k \sqrt{|g(z)||g(z')|
}\delta A_\mu(z,-k)\Pi_{CG,EC}^{\mu\nu}(z,z',k) \delta A_\nu(z',k)\,
\eeqn
where $\Pi^{\mu\nu}_{CG,EC}$ is a radially coarse grained polarization tensor.  
To obtain an approximate value for $\Pi^{\mu\nu}_{CG,EC}$, we start from the 
polarization tensor for relativistic fermions in each local Lorentz frame, $\Pi^{\mu\nu}.$  
Inside the fluid, we will only be interested in the behavior of the polarization tensor at 
long wavelengths in the radial direction. We define a coarse grained polarization 
tensor in flat space by simply projecting on to infinite wavelength,
\beqn
\Pi^{\mu\nu}_{CG,flat}(k_z,k_{L.L.})&\equiv&\Pi^{\mu\nu}(0,k_{L.L.})\,,\\
\rightarrow \Pi^{\mu\nu}_{CG,flat}(z,z',k_{L.L.}) &=& \delta(z-z')\Pi^{\mu\nu}(0,k_{L.L.})\,,\eeqn
where the spatial momentum in each local frame is denoted $k_{L.L.}$.

Locally, we then approximate the (coarse grained) curved space polarization tensor 
by pulling back the (coarse grained) flat space expression, using the local values of the chemical potential and Fermi momentum.  This last step introduces a subtlety by 
redshifting the spatial momentum which is to be used in the polarization tensor.  
Because the fluid in each patch naturally lives in the local Lorentz frame, the 
momentum flowing through the polarization tensor is that of the local Lorentz 
frame, $k_{L.L}=z k.$  Pulling all this together, we arrive at the polarization 
contribution to the effective action
\beqn 
S_{Pol}= \frac{e^2}{2}\int dz \sqrt{|g|}\delta A_\mu(z,-k)\Pi^{\mu\nu}(0,z k)
\delta A_\nu(z,k)\,,
\eeqn
written in terms of the flat space polarization tensor evaluated in each fluid patch. 
The full expressions are quite lengthy and may be found in Appendix~\ref{ap:FlatSpPol}.

\subsubsection{Lack of Friedel Oscillations}

As in Section~\ref{sec:NRpol}, we are only interested in the additional polarization due 
to the finite density of fermionic matter.  For brevity we will suppress all contributions not 
relevant for triggering Friedel oscillations. In each local patch, relativistic Friedel oscillations 
are caused by a mildly singular term (see Appendix~\ref{ap:FlatSpPol} for details), 
\beq \Pi^{\mu\nu}(0,k_{L.L.}) 
\sim N^{\mu\nu}(k_{L.L.})\ln\left(\frac{k_{L.L.}-2k_f}{k_{L.L.}+2 k_f}\right)+... \,,\eeq
where $N^{\mu\nu}$ has smooth dependence on the momentum near $k_f$ and the 
neglected terms are also smooth at $k_f$. With this, the relevant term in the gauge 
field effective action becomes
\beqn 
S_{Pol}\sim \frac{e^2}{2}\int dz \sqrt{|g|}\delta A_\mu(z,{-}k)\left[N^{\mu\nu}(z k)
\ln\left(\frac{z k-2\sqrt{\frac{(z h(z))^2}{f(z)}-\hat{m}^2 }}{z k
+2\sqrt{\frac{(z h(z))^2}{f(z)}-\hat{m}^2 }}\right)\right] \delta A_\nu(z,k)\,. \nonumber\\
\eeqn
There is an important difference between this polarization tensor and that of 
Section~\ref{sec:BoxVacPol}.  In the hard wall example, the source of the bulk 
Friedel oscillations was $z$ independent.  This led to a factorization of radial 
and boundary information such that the bulk Friedel oscillations were coherent 
across the radial direction.  For the electron star, the bulk Fermi momentum is 
a local concept.  Even in the deep interior of an electron star, where the local 
chemical potential goes to a constant, the polarization is $z$ dependent,
\beqn 
S_{Pol}\sim \frac{e^2}{2}\int_{\mathrm{large}\ z} dz 
\sqrt{|g|}\delta A_\mu(z,-k)\left[N^{\mu\nu}(z k)\ln\left(\frac{z k
-2\sqrt{h_\infty^2-\hat{m}^2}}{z k+2\sqrt{h_\infty^2-\hat{m}^2}}\right)\right]\delta A_\nu(z,k)\,.\nonumber\\
\eeqn
We see that different bulk oscillations are triggered at each radius.

The absence of boundary Friedel oscillations in electron stars has two basic causes:
there is a continuum of bulk Fermi surfaces and each surface is localized in the bulk.  
These facts are manifest in the smearing in two ways.  First, the argument of the 
polarization is a function of the background profiles for the metric and gauge field. 
Given this, one can expect a degree of smoothing set by the sharpness of the 
transition from the interior Lifshitz to the asymptotic $\ads$ behavior.  The second, 
and more important, manifestation is the momentum redshift factor which persists all 
the way into the Lifshitz scaling regime. This is analogous to the smearing 
described in \cite{Hartnoll:2010xj}.  
Since the fate of a continuum of bulk Fermi surfaces may be of relevance for holographic model building,
it is important to understand to what extent the two sources of smearing we have identified in this work are independent. 

One might complain that the polarization effects were included in an {\it ad hoc} 
manner in this section.  However, the approximations we have made are more or 
less optimal for seeing Friedel oscillations and yet there is still smearing due to the 
infinite number of bulk Fermi surfaces.  A more thorough analysis would keep track 
of the radial scale over which the fermions are integrated out, almost certainly leading 
to even less sharp features than seen with the simple projection used here.


\section{Discussion \label{sec:Discussion}}

In this paper, we have initiated a study of Fermi surfaces in holographic metallic 
states, without reference to probe fermions.  Specifically, we use static current-current 
correlation functions to characterize boundary Fermi surfaces, focusing on the question 
of when bulk Friedel oscillations trigger boundary Friedel oscillations.  We discussed 
two systems of charged bulk matter, not hidden behind horizons: fermions in 
a hard wall truncation of $\ads$ and an electron star geometry.  For both of these 
systems a Luttinger count indicates that the dual theory has filled Fermi surfaces 
and for both systems it is necessary to consider loop effects in the current-current 
susceptibilities to see potential oscillations.

Fermions in the confining model geometry clearly display bulk Friedel oscillations.  
Because confinement leads to discrete bulk Fermi surfaces and each bulk Fermi 
surface is non-local in the radial direction, the bulk oscillations map into boundary 
Friedel oscillations.  The loop parameter, ``$1/N$," sets the amplitude of the boundary 
Friedel oscillations, but not the wave vector at which they occur.

For the electron star, on the other hand, the bulk Fermi surfaces are local features 
which vary along the radial direction.  We find that the local character of the bulk Fermi 
surface prevents bulk Friedel oscillations from combining coherently. Each radial 
position sources a Friedel oscillation at a different boundary momentum.  
This is consistent with a boundary theory comprised of a continuum of Fermi 
surfaces, as proposed in \cite{Hartnoll:2010xj}.

Extrapolating from these two examples a stronger claim would be: For holographic 
metallic states to exhibit a discrete set of boundary Fermi surfaces in all channels, 
there must be a discrete set of bulk Fermi surfaces and each bulk Fermi surfaces 
must be completely delocalized in the bulk.  If either of these conditions fails any 
sharp features will be washed out. 

In this paper we only analyze systems with either completely delocalized bulk Fermi 
surfaces or completely localized surfaces. Testing whether the lessons learned from our 
two examples are general is of obvious importance. An alternative system is provided 
by the $\ads$ black hole with Dirac hair \cite{Cubrovic:2010bf} and it would be 
interesting to obtain the static susceptibility for this case as well.
It would also be interesting to repeat the analysis of fermions in a hard wall geometry 
with alternate boundary conditions.  As discussed in \cite{Sachdev:2011ze}, these 
should be dual to non-Fermi liquids and it would be instructive to see how this simple 
change causes such drastic effects in the boundary theory.  Finally, it would be 
instructive to incorporate the ideas of \cite{Sachdev:2011ze} in a more realistic 
model of confinement \cite{Witten:1998zw} and to see the effect of Fermi surfaces 
on other transport quantities.


\acknowledgments

We would like to thank E. J. Brynjolfsson, J. Hertz, N. Jokela, E. Keski-Vakkuri, 
S. Sachdev, K. Schalm, K. P. Yogendran, and K. Zarembo for helpful discussions. 
This work was supported in part by the Icelandic Research Fund and by the 
University of Iceland Research Fund.


\begin{appendix}


\section{Perturbative Polarization Tensor\label{ap:NRPol}}

In Section~\ref{sec:NRpol} we summarized the perturbative computation 
of the non-relativistic static density-density polarization,  
\beq
\Pi(x\, t,x'\, t') = -ie^2\mathcal{G}(x\, t,x'\, t')\mathcal{G}(x'\, t',x\, t)\,.
\eeq 
This is textbook material (see \cite{Fetter} for example) and details are only presented in this Appendix to make the paper more self-contained.  

As usual, it is simplest to 
analyze the polarization in momentum space, 
\beq
\Pi(q,\nu) = -ie^2\int  \frac{d\omega d\vec{k}}{(2\pi)^{D+1}}
\mathcal{G} (k,\omega)\mathcal{G} (k+q,\omega+\nu).
\eeq
Looking at the denominators of the time ordered Green's function \eqref{green_function_ex} shows that only when $k$ and $k+q$ lie on opposite sides of $k_f$ will the answer be non-zero,
\beqn
\Pi(q,\nu) 
&=& e^2 \int \frac{d\vec{k}}{(2\pi)^D}
\left[\frac{\theta(|q+k|-k_f)\theta(k_f-|k|)}{\nu-(E_{k+q}-E_{k})+i\eta}-\frac{\theta(|k|-k_f)\theta(k_f-|q+k|)}{\nu-(E_{k+q}-E_{k})-i\eta}\right]\,\nonumber\\\nonumber\\
&=&e^2 \int \frac{d\vec{k}}{(2\pi)^D}\theta(|q+k|-k_f)\theta(k_f-|k|)\nonumber\\&&\qquad\times\left(\frac{1}{\nu-(E_{k+q}-E_{k})+i\eta}-\frac{1}{\nu+(E_{k+q}-E_{k})-i\eta}\right)\,. 
\eeqn
In the second term we have changed momentum variables $k\rightarrow -k-q$ and assumed that the spectrum is invariant under sign changes (that the Fermi surface is spherical).
The next step is to extract the imaginary part%
\footnote{Recall that for real $\omega$ \beq \frac{1}{\omega-\omega'\pm i\eta}=\mathcal{P}\frac{1}{\omega-\omega'}\mp\pi i \delta(\omega-\omega')\,.\eeq}
\beqn
\mathrm{Im}\ \Pi(q,\nu)
&=&  - \pi e^2 \int \frac{d\vec{k}}{(2\pi)^D}\theta(|q+k|-k_f)\theta(k_f-|k|)\nonumber\\&&\ \ \times\left[\delta\left(\nu-(E_{k+q}-E_{k})\right)+\delta\left(\nu+(E_{k+q}-E_{k})\right)\right]\,. \nn
\eeqn

\subsection{Static Limit \label{ap:NREx}}
If we let $\nu\rightarrow0$ with $q$ fixed, we immediately see 
\beq
\mathrm{Im}\ \Pi(q,0)=0\,.
\eeq  There are two Fermi spheres in the problem.  The step functions force you to be inside one and outside the other.  On the other hand, the delta functions force the Fermi spheres to coincide and there is no solution to the constraints.

From the static real part we have
\beqn
\mathrm{Re}\, \Pi(q,0) 
&=&  -e^2\mathcal{P}\int \frac{d\vec{k}}{(2\pi)^D}\left[1-\theta(k_f-|k+q|)\right]\theta(k_f-|k|)\frac{2}{(E_{k+q}-E_{k})}\,\nonumber\\\nonumber\\
&=& -e^2\mathcal{P}\int \frac{d\vec{k}}{(2\pi)^D}\theta(k_f-|k|)\frac{2}{(E_{k+q}-E_{k})}\,.
\eeqn
Here we have used antisymmetry under $k\leftrightarrow k+q$.  In $2+1$ dimensions we have
\beqn
\mathrm{Re}\, \Pi(q,0) =-\frac{e^2}{2\pi^2}\mathcal{P}\int d k d\theta\ \theta(k_f-k)\frac{k}{E_{k+q}-E_{k}}\,.
\eeqn
For spherical Fermi surfaces the integrand is singular when $q=2k_f$.  In general, the polarization tensor has non-analytic behavior in the momentum space and oscillatory behavior in position space.

For example, for non-relativistic fermions we have $E_{k+q}-E_k=q^2/2m+k q \cos(\theta)/m.$  The polarization becomes,  
\beqn
\mathrm{Re} \, \Pi(q,0) 
&=& -\frac{me^2}{2 \pi^2 q}\mathcal{P}\int d k d\theta\ \theta(k_f-k)\frac{k}{q/2+k\cos(\theta)}\,\\
&=&-\frac{me^2}{2\pi}\left(1-\sqrt{1-\left(\frac{2k_f}{q}\right)^2}\theta(q-2k_f)\right)\,.
\eeqn


\section{Holographic Relative Polarization Tensor\label{ap:HolRelPol}}

In Section~\ref{sec:NRpol} we introduced the relative polarization tensor,
\beqn
\Pi^{\mu\nu}_{rel}(z,z',\vec{p})
&=& -\int \frac{d\omega \, d^{2}k}{(2\pi)^3}  \mathrm{Tr}\ \Big{(}M^\mu G(\omega,k,z,z')M^\nu  G(\omega ,k+p,z',z)\Big{)}\nonumber\\
&&\qquad \qquad -  \mathrm{Tr}\ \Big{(}M^\mu G^0(\omega,k,z,z')M^\nu  G^0(\omega ,k+p,z',z)\Big{)}\,.
\eeqn
 Using Eq. (\ref{eq:FermGF}) we can perform the frequency integral:
\beqn\label{eq:FullRelPol}
&&\Pi_{rel}^{\mu\nu}(z,z',\vec{p})=\nn \\
&& \sum_{\ell\ell'}\,' \int \frac{ d^{2}k}{(2\pi)^2}\left[\frac{\theta(E_{\ell'}(k+p))\theta(-E_{\ell}(k))}{E_{\ell'}(k+p)-E_{\ell}(k)+i\eta\mathrm{s}(E_{\ell}(k))}
-\frac{\theta(-E_{\ell'}(k+p))\theta(E_{\ell}(k))}{E_{\ell'}(k+p)-E_{\ell}(k)-i\eta\mathrm{s}(E_{\ell}(k))}\right]\nn\\
&&   \quad    \times  \mathrm{Tr}\ M^\mu\ \chi_{\ell,k}(z)\chi_{\ell,k}^\dagger(z')M^\nu\ \chi_{\ell',k+p}(z')\chi_{\ell',k+p}^\dagger(z)\nonumber\\
&&- \int \frac{ d^{2}k}{(2\pi)^2}\left[\frac{\theta(\ell')\theta(-\ell)}{E^0_{\ell',k+p}-E^0_{\ell,k}-i\eta}-\frac{\theta(-\ell')\theta(\ell)}{E^0_{\ell',k+p}-E^0_{\ell,k}-i\eta}\right]\nonumber\\
&&\quad \times  \mathrm{Tr}\  M^\mu\ \chi^0_{\ell,k}(z)\chi^{0\dagger}_{\ell,k}(z')M^\nu\ \chi^0_{\ell',k+p}(z')\chi^{0\dagger}_{\ell',k+p}(z)\,.
\eeqn
In this we have denoted the contributions with zero chemical potential with  ${}^0$ and the primed summation indicates that the mode sums have no $\ell,\ell' = 0$ terms.  Also note that we have introduced the notation $s(x)\equiv \mathrm{sign}(x)$.

Though expression in Eq. (\ref{eq:FullRelPol}) is quite complicated, there are two important features to note.  Firstly, it is finite in the bulk UV.  The large momentum and radial mode contributions do not know about the low energy finite density physics.  Secondly, spherical bulk Fermi surfaces imply singular behavior due to the integrand's denominator at $p=2 k_f$ and $\ell=\ell'$.  The only terms where $\ell=\ell'$ are when $k$ and $k+p$ lie on opposite sides of a Fermi surface.  Quite generally, these singularities in the integrand will source Friedel oscillations in position space.  

\subsection{High Confinement Scale and Non-Relativistic Limit\label{ap:Simplification}}
We will now  consider a simplifying situation where there is a single Fermi surface occurring at $k=k_f$ and $\ell=1$.  In this case we may split the summand in Eq. (\ref{eq:FullRelPol}) into two classes, $\ell=\ell'=1$ and all other combinations,
\be
&& \Pi_{rel}^{\mu\nu}(z,z',\vec{p}) =\nn\\
&&  \int \frac{ d^{2}k}{(2\pi)^2}\left(\frac{\theta(|k+p|-k_f)\theta(k_f-|k|)}{E_{1}(k+p)-E_{1}(k)+i\eta\mathrm{s}(E_{1}(k))}-\frac{\theta(k_f-|k+p|)\theta(|k|-k_f)}{E_{1}(k+p)-E_{1}(k)-i\eta\mathrm{s}(E_{1}(k))}\right)\nn\\
&& \qquad \times  \mathrm{Tr}\ \,M^\mu\, \chi_{1,k}(z)\chi_{1,k}^\dagger(z')\, M^\nu\, \chi_{1,k+p}(z')\chi_{1,k+p}^\dagger(z)
+\Pi^{\mu\nu}_{analytic}\label{eq:AdSBoxPol}\,.
\ee
If $\ell=\ell'=1$ the integrand has a mild singularity due to the Fermi surface and there is no contribution from the vacuum terms.  The rest of the mode number sums are analytic in the external momenta and do not involve the Fermi surface. 

It is difficult to deal with the analytic contributions without explicit expressions for the energy eigenvalues and wave functions.  However, we do note that their dominant contributions come from small momenta where the energy values scale with $E_\ell(0)\sim\frac{1}{z_m}.$  Consequently, the analytic contributions to the relative vacuum polarization scale as $z_m$.  If we model a high confinement scale, we can safely ignore all but the $\ell=\ell'=1$ contributions,
\be
&& \Pi_{rel}^{\mu\nu}(z,z',\vec{p}) =\nn\\
&&   \int \frac{ d^{2}k}{(2\pi)^2}\left(\frac{\theta(|k+p|-k_f)\theta(k_f-|k|)}{E_{1}(k+p)-E_{1}(k)+i\eta\mathrm{s}(E_{1}(k))}-\frac{\theta(k_f-|k+p|)\theta(|k|-k_f)}{E_{1}(k+p)-E_{1}(k)-i\eta\mathrm{s}(E_{1}(k))}\right)\nonumber\\
&&\qquad \times  \mathrm{Tr}\ M^\mu\ \chi_{1,k}(z)\chi_{1,k}^\dagger(z')\, M^\nu\ \chi_{1,k+p}(z')\chi_{1,k+p}^\dagger(z) +\mathcal{O}(z_m)\,.
\ee 

While this is a great simplification, it is not quite enough for us to be able to find an analytic expression.  In order to go further we will take a non-relativistic limit and scale the chemical potential towards the bulk mass.  In this way we are left with a small $k_f$ and numerically observe that, for small momenta, the energy spectrum is well approximated by,
\beq
E_1(k)\sim a_1+b_1 k^2\,.
\eeq
Using this functional form the holographic calculation closely parallels the $2+1$ non-relativistic example in Section~\ref{ap:NREx} with $2 b_1$ playing the role of an inverse mass,
\be
&& \Pi_{rel}^{\mu\nu}(z,z',\vec{p})  =\nn\\
&&
\frac{1}{p b_1} \int \frac{ d^{2}k}{(2\pi)^2}\Big[ \frac{\theta(E_{1}(k+p))\theta(-E_{1}(k))}{p+2\cos(\theta)k+i\eta\mathrm{s}(E_1(k))} 
\nn\\
&&\qquad\times \mathrm{Tr} \left(M^\mu\ \chi_{1,k}(z)\chi_{1,k}^\dagger(z')M^\nu\ \chi_{1,k+p}(z')\chi_{1,k+p}^\dagger(z)\right)  \nonumber\\
&&-  \frac{\theta(E_{1 }(k-p))\theta(-E_{1}(k))}{ p-2\cos(\theta)k-i\eta\mathrm{s}(E_{1}(k))}
 \mathrm{Tr} \left(M^\mu\ \chi_{1,k-p}(z)\chi_{1,k-p}^\dagger(z')M^\nu\ \chi_{1,k }(z')\chi_{1,k }^\dagger(z)\right)\Big]\,.\nonumber
\\
\eeqn
As in the non-relativistic case, the static polarization is purely real.  We have 
\beqn
\Pi_{rel}^{\mu\nu}(z,z',\vec{p})
&=& \frac{1}{4 b_1 \pi^2 p}\mathcal{P}\int d k d\theta\ \theta(k_f-k)\frac{k}{p/2+k\cos(\theta)}\nn\\
&& \quad\times  \mathrm{Tr} \left(M^\mu\ \chi_{1,k }(z)\chi_{1,k }^\dagger(z')M^\nu\ \chi_{1,k+p }(z')\chi_{1,k +p}^\dagger(z)\right)\,.\label{eq:NRVPol}
 \eeqn
Numerically, it is expensive to work with Eq. (\ref{eq:NRVPol}) without further approximation.  We can use the fact that $k_f$ is small which means that the magnitude of the loop momentum is never large and that the wave functions are slowly varying functions of the momenta.  Therefore we may expand the wave function factors for small loop momenta, $k$,
\be
&&\mathrm{Tr} \left(M^0\ \chi_{1,k }(z)\chi_{1,k }^\dagger(z')M^0\ \chi_{1,k+p}(z')\chi_{1,k +p}^\dagger(z)\right) 
\nn\\ 
&&\sim  \mathrm{Tr} \left(M^0\ \chi_{1,0 }(z)\chi_{1,0 }^\dagger(z')M^0\ \chi_{1,p }(z')\chi_{1, p}^\dagger(z)\right)\,.
\ee
This approximation loses some of the angular information in the wavefunctions, but is sufficient to display the essential features.   In this way we find,
\beqn
\Pi_{rel}^{\mu\nu}(z,z',\vec{p}) 
&=&
  \frac{1}{4 b_1 \pi^2 p} \mathrm{Tr} \left(M^\mu\ \chi_{1,0 }(z)\chi_{1,0 }^\dagger(z')M^\nu\ \chi_{1, p }(z')\chi_{1, p}^\dagger(z)\right)\nonumber\\&&
\times\mathcal{P}\int d k d\theta\ \theta(k_f-k)\frac{k}{p/2+k\cos(\theta)} \,.
\eeqn
Introducing the effective expansion parameter, $\lambda  \equiv \frac{1}{4\pi b_1}$,
\beqn \Pi_{rel}^{\mu\nu}(z,z',\vec{p})&=&- \lambda \mathrm{Tr} \left(M^\mu\ \chi_{1,0 }(z)\chi_{1,0 }^\dagger(z')M^\nu\ \chi_{1, p }(z')\chi_{1, p}^\dagger(z)\right) \nonumber\\
&&\times\left(1-\sqrt{1-\left(\frac{2k_f}{p}\right)^2}\theta(p-2k_f)\right)\,.
\eeqn
This is the form of the polarization used in Section~\ref{sec:Numerics}.


\section{Flat Space QED Polarization Tensor\label{ap:FlatSpPol}}
We can understand the emergence of Friedel oscillations in relativistic Fermi liquid theory by studying how weakly coupled relativistic fermions respond to changes in an external gauge field.  When working at finite chemical potential, the Lorentz invariance is broken and the polarization tensor may only be decomposed using the residual spatial rotation symmetry.\footnote{For a general reference to relativistic field theories at finite temperature/density see \cite{Kapusta:2006pm}.}  If we introduce projections onto spatial directions, $P_L$, and the transverse direction, $P_T$, we may write \beq \Pi^{\mu\nu} = G P_T^{\mu\nu}+F P_L^{\mu\nu}\,.\eeq  Expressing the projectors in terms of the spatial momenta, we have \beqn P_T^{ij} &=& \delta^{ij}-\frac{\vec{k}^i\vec{k}^j}{\vec{k}^2}\ \ \ \ \ \ \ \ (\mathrm{only\ spatial\ directions}\neq 0)\,,\\
P_L^{\mu\nu} &=& k^{\mu}k^{\nu}/k^2-g^{\mu\nu}-P_L^{\mu\nu}\,.\eeqn

As is mentioned in Section \ref{sec:BoxVacPol} in finite density computations, it is useful to further split the polarization tensor into a ``vacuum" part and the relative, finite density, contribution,\beqn \Pi^{\mu\nu}&=&\Pi^{\mu\nu}_{vac}+\Pi^{\mu\nu}_{rel}\,,\\
\Pi^{\mu\nu}_{vac}&=& \Pi^{\mu\nu}|_{\mu,T\rightarrow 0}\,.\eeqn  
The vacuum contribution is a standard textbook computation \cite{Kapusta:2006pm}.    Focusing on $3+1$ dimensions, $F(0,\vec{k})$, may be found in \cite{Kapusta:1988fi,Kapusta:2006pm},
\beqn \left(\frac{24 \pi^2}{e^2}\right) F_{rel}(0,k) &=& 16 \mu k_f + 4 k^2 \ln\left(\frac{m}{\mu + k_f}\right)  \nonumber \\
&& + \left(\frac{\mu}{k}\right)\left(3 k^2-4 \mu^2\right)\ln\left(\frac{k-2k_f}{k+2k_f}\right)^2  \\ 
&& + \left(\frac{2m^2-k^2}{k}\right)\sqrt{k^2+4m^2}\ln\left(\frac{\sqrt{k^2+4m^2} k_f-k\mu}{\sqrt{k^2+4m^2} k_f-k\mu}\right)^2\,.\nonumber \eeqn 
A similar straightforward calculation yields 
\beqn 
\left(\frac{24 \pi^2}{e^2}\right) G_{rel}(0,k) 
&=& 4 \mu k_f - 8 k^2\ln\left( \frac{m}{\mu+k_f} \right)  \nonumber\\ &&+  \mu \left(\frac{ 4 \mu^2-12 m^2 - 9 k^2}{2k}\right)\ln\left(\frac{k-2k_f}{k+2k_f}\right)^2  \\ &&-\frac{(4m^2+k^2)^{3/2}}{k}\ln\left(\frac{\sqrt{k^2+4m^2} k_f-\mu k}{\sqrt{k^2+4m^2} k_f+ \mu k}\right)^2 \,.\nonumber \eeqn
The vacuum contributions contain the usual ultraviolet divergences for QED, and contain no information about the Fermi surface which is set by $\mu$.  For us, the most important feature is the logarithmic term with a singularity at $k=2k_f$ in both of these functions.  Upon Fourier transformation this gives rise to spatial variations with wave number $2k_f$.  
When using this in the holographic calculation one should note that the appropriate values of $k$ and $k_f$ are functions of the radial coordinate through redshift factors.


\end{appendix}


\addcontentsline{toc}{section}{References}



\providecommand{\href}[2]{#2}\begingroup\raggedright\endgroup

\end{document}